\documentclass[sigconf]{acmart}

\makeatother
\usepackage{amsthm}
\usepackage{enumitem}
\usepackage{subcaption}
\usepackage{tikz}
\usetikzlibrary{matrix}
\usepackage{tikz}
\usepackage{pgfplots} 
\pgfplotsset{compat=1.16} 
\usepackage{balance}
\usepackage{bigstrut,array,multirow,tabularx}
\usepackage{caption}
\usepackage{amssymb}
\usepackage{pifont}
\newcommand{\cmark}{\ding{51}}%
\newcommand{\xmark}{\ding{55}}%
\usepackage{dsfont}
\usepackage{mathtools}

\usepackage{algorithm}
\usepackage{algorithmic}
\usepackage{makecell} 
\usepackage{booktabs}
\usepackage{graphicx}
\usepackage{pgfplots}
\pgfplotsset{compat=1.18}
\usepackage{subcaption}
\definecolor{lightblue}{RGB}{0.93,0.95,1.0} 
\definecolor{lightred}{RGB}{1.0,0.93,0.93} 
\usepackage{pgfplots}
\usepackage{pgfplotstable}
\usepackage{colortbl}
\usepackage{booktabs}
\usepackage{array}
\usepackage{filecontents}
\usepackage{pifont}
\usepackage[export]{adjustbox}

\AtBeginDocument{%
  }

\copyrightyear{2025}
\acmYear{2025}
\setcopyright{cc}
\setcctype{by}
\acmConference[WWW Companion '25]{Companion Proceedings of the ACM Web Conference 2025}{April 28-May 2, 2025}{Sydney, NSW, Australia}
\acmBooktitle{Companion Proceedings of the ACM Web Conference 2025 (WWW Companion '25), April 28-May 2, 2025, Sydney, NSW, Australia}
\acmDOI{10.1145/3701716.3715482}
\acmISBN{979-8-4007-1331-6/25/04}

\makeatletter
\gdef\@copyrightpermission{
 \begin{minipage}{0.2\columnwidth}
  \href{https://creativecommons.org/licenses/by/4.0/}
  {\includegraphics[width=0.90\textwidth]{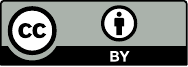}}
 \end{minipage}\hfill
 \begin{minipage}{0.8\columnwidth}
  \href{https://creativecommons.org/licenses/by/4.0/}{This work is licensed under a Creative Commons Attribution International 4.0 License.}
 \end{minipage}
 \vspace{5pt}
}
\makeatother

\begin{document}

\title[Leveraging Member–Group Relations via Multi-View Graph Filtering 
       \protect\linebreak for Effective Group Recommendation]{
       Leveraging Member–Group Relations via Multi-View Graph Filtering for Effective Group Recommendation}

\author{Chae-Hyun Kim}
\authornote{Equal contribution (co-first authors)}
\affiliation{%
  \institution{Yonsei University}
  \city{Seoul}
  \country{Republic of Korea}
}
\email{kimchaehyun0315@yonsei.ac.kr}

\author{Yoon-Ryung Choi}
\authornotemark[1]
\affiliation{%
  \institution{Sookmyung Women's University}
  \city{Seoul}
  \country{Republic of Korea}
}
\email{wendych@sookmyung.ac.kr}

\author{Jin-Duk Park}
\authornote{Co-corresponding authors}
\affiliation{%
  \institution{Yonsei University}
  \city{Seoul}
  \country{Republic of Korea}
}
\email{jindeok6@yonsei.ac.kr}

\author{Won-Yong Shin}
\authornotemark[2]
\affiliation{%
  \institution{Yonsei University}
  \city{Seoul}
  \country{Republic of Korea}
}
\email{wy.shin@yonsei.ac.kr}

\renewcommand{\shortauthors}{Chae-Hyun Kim, Yoon-Ryung Choi, Jin-Duk Park, and Won-Yong Shin}

\begin{abstract}
 Group recommendation aims at providing optimized recommendations tailored to diverse groups, enabling groups to enjoy appropriate items. On the other hand, most existing group recommendation methods are built upon deep neural network (DNN) architectures designed to capture the intricate relationships between member-level and group-level interactions. While these DNN-based approaches have proven their effectiveness, they require complex and expensive training procedures to incorporate group-level interactions in addition to member-level interactions. To overcome such limitations, we introduce \textsf{Group-GF}, a new approach for extremely fast recommendations of items to each group via \textit{multi-view graph filtering (GF)} that offers a holistic view of complex member--group dynamics, without the need for costly model training. Specifically, in \textsf{Group-GF}, we first construct three item similarity graphs manifesting different viewpoints for GF. Then, we discover a {\it distinct} polynomial graph filter for each similarity graph and judiciously aggregate the three graph filters. Extensive experiments demonstrate the effectiveness of \textsf{Group-GF} in terms of significantly reducing runtime and achieving state-of-the-art recommendation accuracy. 
\end{abstract}



\maketitle

\section{Introduction}
\label{section 1} 

\begin{figure}[t]
    \centering
    \includegraphics[width=0.9\columnwidth]{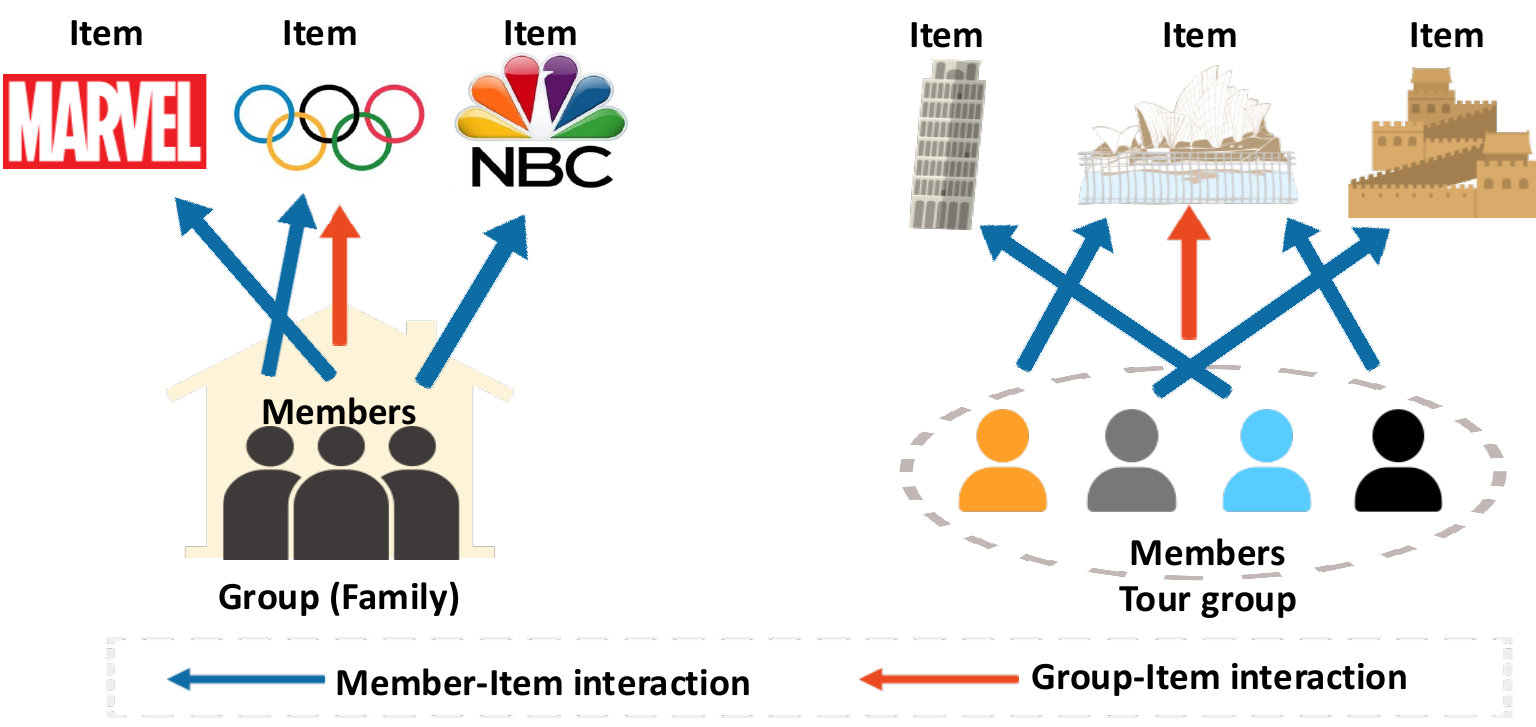}
    \caption{Examples of both group and member consumption patterns on TV and trip platforms.}
    \label{fig_intro}
\end{figure}

Group recommender systems aim to provide precise recommendations tailored to a collection of members rather than individuals \cite{guo2021hierarchical, cao2018attentive, jia2021hypergraph, wu2023consrec, sankar2020groupim}. In group recommendations, capturing the intricate relationships between member-level and group-level interactions is essential for accurate group recommendations \cite{cao2018attentive, wu2023consrec}. For example, the preferences of a travel group on a trip platform may differ from individual members' preferences, as shown in Figure \ref{fig_intro}. In this context, it is of paramount importance to effectively integrate such dissimilar preferences for accurate group recommendations \cite{hu2014deep, baltrunas2010group}. In other words, treating both member-level and group-level interactions separately can result in a failure to account for the nuanced dynamics that occur within group contexts \cite{yadati2019hypergcn, yang2023group, jia2021hypergraph}. 

To jointly deal with both member-level and group-level interactions, existing approaches often resort to hypergraph-based \cite{yadati2019hypergcn, yang2023group, wu2023consrec, jia2021hypergraph} and self-supervised learning (SSL)-based \cite{zhang2021double,wu2023consrec, sankar2020groupim} methods. However, the aforementioned methods often involve complex hypergraph modeling or expensive model training costs for SSL, which can hinder their responsiveness to rapidly changing member preferences.  

To address these practical challenges, we propose \textsf{Group-GF}, the first attempt at group recommendations built upon {\it graph filtering} {\it (GF)}. Our \textsf{Group-GF} method is composed of 1) the construction of three item similarity graphs exhibiting different viewpoints and 2) the optimal design of {\it distinct} polynomial graph filters that are hardware-friendly without costly matrix decomposition. More precisely, \textsf{Group-GF} starts by constructing two {\it augmented item similarity graphs} concatenating the member--group matrix and a {\it unified item similarity graph} concatenating the  member--item and group--item interaction matrices, thereby seamlessly integrating member and group information for accurate group recommendations. Next, based on the three constructed graphs, we {\it distinctly} and {\it optimally} perform polynomial GF for each similarity graph and then aggregate the three graph filters.

Through extensive evaluations on benchmark datasets, \textsf{Group-GF} achieves not only state-of-the-art accuracy but also extraordinary \textbf{efficient runtime up to 1.55 seconds}. This remarkable performance can be attributed to its training-free multi-view GF, which accommodates complex member--group dynamics while relying solely on simple matrix operations. In addition, we theoretically connect \textsf{Group-GF}’s filtering process to optimization with smoothness regularization, offering clearer interpretability of the model’s behavior.

\section{Preliminary}
\label{section 2}
\subsection{Graph Filtering}
We provide fundamental principles of GF. Suppose a weighted graph $G = (V, E)$ represented by an adjacency matrix $A \in\mathbb{R}^{|V|\times |V|}$. A $d$-dimensional vector $\mathbf{x} \in \mathbb{R}^{|V|}$ is a graph signal, where $x_i$ represents the signal strength of node $i$ in ${\bf x}$. 
The graph Laplacian $L$ of $G$ is $L = D - A$, where $D=\text{diag}(A\mathbf{1})$. 
The smoothness measure $S({\bf x})$ is expressed as \cite{shen2021powerful, shuman2013emerging}: 
    \begin{equation}
        \label{eq:Sm_1}
        S({\bf x}) = \sum_{i,j}A_{i,j}(x_i-x_j)^2 =  \mathbf{x}^T L \mathbf{x}.
    \end{equation} 
Smaller values of $S(x)$ indicate that the signal $\mathbf{x}$ is smoother on the graph. By the eigen-decomposition $L = U\Lambda U^T$, we can formally define the graph Fourier transform (GFT) of a graph signal $\mathbf{x}$ as $\hat{\mathbf{x}} = U^T \mathbf{x}$, where $U\in\mathbb{R}^{|V|\times |V|}$ is the matrix whose $i$-th column is the eigenvector $u_i$ of $L$. Here, the signal $\mathbf{x}$ is considered smooth if the dot product of the eigenvectors corresponding to smaller eigenvalues of $L$ is high. Given a graph Laplacian matrix $L$, a graph filter $H(L)\in\mathbb{R}^{|V|\times |V|}$ is given by
\begin{equation}
H(L) = U \text{diag}(h(\lambda_1), \cdots, h(\lambda_{|V|})) U^T,
\end{equation}
where $h:\mathbb{C} \rightarrow \mathbb{R}$ is the frequency response function that maps eigenvalues $\{\lambda_1,\cdots,\lambda_{|V|}\}$ of $L$ to $\{h(\lambda_1),\cdots,h(\lambda_{|V|})\}$. The convolution of a  graph signal $\mathbf{x}$ with a graph filter $H(L)$ is then represented as $H(L)\mathbf{x}=U\text{diag}(h(\lambda_1), \cdots, h(\lambda_{|V|})) U^T\mathbf{x}$.

\subsection{Problem Definition}
Consider the sets of members, items, and groups, denoted by $\mathcal{U} = \{u_1, u_2, \cdots, u_{|\mathcal{U}|}\}$, $\mathcal{I} = \{i_1, i_2, \cdots, i_{|\mathcal{I}|}\}$, and $\mathcal{G} = \{g_1, g_2, \ldots, g_{|\mathcal{G}|}\}$, respectively. Each member and group interacts with various items, reflecting their own preferences. Among $\mathcal{U}$, $\mathcal{I}$, and $\mathcal{G}$, there are two types of interactions: 1) interactions between members and items and 2) interactions between groups and items.
We denote the member--item interaction matrix by $R_u \in \mathbb{R}^{|\mathcal{U}|\times|\mathcal{I}|}$ and the group--item interaction matrix by $R_g \in \mathbb{R}^{|\mathcal{G}|\times|\mathcal{I}|}$, where $r_{ui}= 1$ ({\it resp.} $r_{gi}= 1) $ if member $u$ ({\it resp.} group $g$) is associated with item $i$, and $r_{ui}= 0$ ({\it resp.} $r_{gi}= 0) $ otherwise. The $t$-th group $\mathcal{G}_t=\{u_1,u_2,\cdots,u_s,\cdots,u_{|\mathcal{G}_t|} \} \in \mathcal{G}$ consists of a set of members in the belonging group. For a given target group $\mathcal{G}_t$, the objective of the group recommendation task is to recommend items to {\it each group} $\mathcal{G}_t$ \cite{wu2023consrec, jia2021hypergraph}.

\begin{figure}[t] 
    \centering 
    \includegraphics[width=0.48\textwidth]{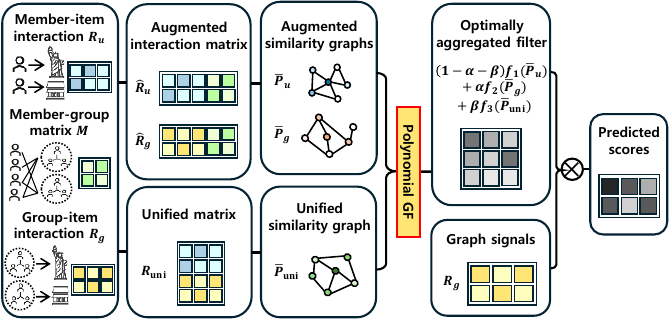} 
    \caption{The schematic overview of \textsf{Group-GF}.} 
    \label{overview}
\end{figure}

\section{Methodology}
In this section, we elaborate on \textsf{Group-GF}, a method that judiciously performs multi-view GF to model complex interactions between members and groups. The schematic overview of \textsf{Group-GF} is shown in Figure \ref{overview}.
\label{sec 3.2}
\subsection{Augmented Graph Construction}\label{sec 3.2.1}
Standard GF-based recommendation methods \cite{liu2023personalized, park2024turbo, shen2021powerful, choi2023blurring, xia2022fire, he2020lightgcn} for the individual recommendation task begin with constructing a graph structure, where each item is represented as a node and the similarities between items are modeled as edges. The construction process is formulated as follows:
\begin{equation}
    \label{canonical_rec_eq}
    \tilde{P} = \tilde{R}^T\tilde{R}; \tilde{R} = D^{-1/2}_{\mathcal{U}}RD^{-1/2}_\mathcal{I}, 
\end{equation}
where the operator | | denotes the concatenation of matrices, and $R \in \mathbb{R}^{|\mathcal{U}| \times |\mathcal{I}|}$ is the member--item interaction matrix; $\tilde{R}$ is the normalized interaction matrix; $D_\mathcal{U}=\text{diag}(R\mathbf{1})$ and $D_\mathcal{I} = \text{diag}(\mathbf{1}^TR)$; and $\tilde{P}$ is the adjacency matrix of the item--item similarity graph. However, constructing separate graphs for member--item and group--item interactions overlooks the member--group relations. Thus, we present a nontrivial strategy that constructs \textit{augmented item similarity graphs} using a member--group relation matrix $M \in \mathbb{R}^{|\mathcal{G}|\times|\mathcal{U}|}$, where $M_{ij} = 1$ indicates that member $j$ belongs to group $i$, and $M_{ij} = 0$ otherwise. We first augment $R_u$ and $R_g$ by concatenating the member--group relation matrix $M$:
\begin{equation}
\label{aug_inters}
    \hat{R}_u = R_u || M^\top;\hat{R}_g = R_g || M,
\end{equation}
where the operator $||$ denotes the concatenation of matrices, and $\hat{R}_u \in \mathbb{R}^{|\mathcal{U}|\times(|\mathcal{I}|+|\mathcal{G}|)}$ and $\hat{R}_g \in \mathbb{R}^{|\mathcal{G}|\times(|\mathcal{I}|+|\mathcal{U}|)}$ are the augmented member-level and group-level interaction matrices, respectively. It is worth noting that this augmentation allows us to jointly leverage the information of member--group relations as well as member-level interactions and group-level interactions for GF. Next, we construct two augmented similarity graphs as follows:
\begin{equation}
\begin{aligned}
    \tilde{P}_u = \tilde{R}_u^\top \tilde{R}_u; \quad \tilde{R}_u = D_\mathcal{U}^{-0.5} \hat{R}_u D_{\mathcal{I}_u}^{-0.5}, \\
    \tilde{P}_g = \tilde{R}_g^\top \tilde{R}_g; \quad \tilde{R}_g = D_\mathcal{G}^{-0.5} \hat{R}_g D_{\mathcal{I}_g}^{-0.5},
    \label{P_tilde}
\end{aligned}   
\end{equation}
where $\tilde{P}_u \in \mathbb{R}^{(|\mathcal{I}|+ |\mathcal{G}|)\times(|\mathcal{I}| + |\mathcal{G}|)}$ and $\tilde{P}_g \in \mathbb{R}^{(|\mathcal{I}| + |\mathcal{U}|)\times(|\mathcal{I}| + |\mathcal{U}|)}$ represent the member-level and group-level item similarity graphs, respectively; $\tilde{R}_u$ and $\tilde{R}_g$ are the normalized augmented interaction matrices; and $D_\mathcal{U} = \text{diag}(\hat{R}_u \mathbf{1})$, $D_{\mathcal{I}_u} = \text{diag}(\mathbf{1}^\top \hat{R}_u)$, $D_\mathcal{G} = \text{diag}(\hat{R}_g \mathbf{1})$, and $D_{\mathcal{I}_g} = \text{diag}(\mathbf{1}^\top \hat{R}_g)$ are used for normalization. To construct item similarity graphs, we use $\tilde{P}_u^{\dagger} = \tilde{P}_{u[:|\mathcal{I}|,\ :|\mathcal{I}|]}$ and $\tilde{P}_g^{\dagger} =\tilde{P}_{g[:|\mathcal{I}|,\ :|\mathcal{I}|]}$ by extracting the first $|\mathcal{I}|$ rows and columns from $\tilde{P}_u$ and $\tilde{P}_g$. Additionally, we adjust the difference of similarities in the constructed item similarity graphs to prevent over/under-smoothing via using the Hadamard power \cite{park2024turbo,park2025criteria}. Finally, we characterize two adjusted similarity graphs $\bar{P}_{u}$ and $\bar{P}_{g} $ as 
\begin{equation}
\label{adj_graph_ug}
    \bar{P}_{u}=\tilde{P}_u^{\mathop{\dagger}^{\circ s}};
    \bar{P}_{g}=\tilde{P}_g^{\mathop{\dagger}^{\circ s}},
\end{equation} 
where $s$ is the adjustment parameter.\footnote{Although using two separate adjustment parameters for $\bar{P}_{u}$ and $\bar{P}_{g}$ certainly yields better results, we use a single hyperparameter $s$ for both graphs for simplicity.}

\subsection{Unified Graph Construction}\label{sec 3.2.2} 
We further construct a \textit{unified item similarity graph} $\bar{P}_{\text{uni}}$ to comprehensively grasp the relationships among items. This unified graph enables us to perceive the extensive preference of items by simultaneously taking into account both member--item and group--item interaction matrices. We begin with concatenating $R_g$ and $R_u$ along the item dimension:
\begin{equation}
    R_{\text{uni}} = \begin{bmatrix} R_g \\ R_u \end{bmatrix} \in \mathbb{R}^{(|\mathcal{G}| + |\mathcal{U}|) \times |\mathcal{I}|},
\end{equation}
which effectively combines all possible interactions at the item level. Next, we normalize $R_{\text{uni}}$ to obtain $\tilde{R}_{\text{uni}}$, then compute the graph as: 
\begin{equation}
\tilde{P}_{\text{uni}} = \tilde{R}_{\text{uni}}^\top \tilde{R}_{\text{uni}}; \quad \tilde{R}_{\text{uni}} = D_{\text{uni}}^{-0.5} R_{\text{uni}} D_{\mathcal{I}_{\text{uni}}}^{-0.5},
\label{p_tilde_uni}
\end{equation}
where $D_{\text{uni}} = \operatorname{diag}(R_{\text{uni}} \mathbf{1})$ and $\quad D_{\mathcal{I}_{\text{uni}}} = \operatorname{diag}(\mathbf{1}^\top R_{\text{uni}})$. Finally,  we also adjust the unified item similarity graph using the Hadamard power as follows:
\begin{equation}
\label{adj_graph_uni}
\bar{P}_{\text{uni}} = \tilde{P}_{\text{uni}}^{\circ s}.
\end{equation}
This process results in the unified item similarity graph $\bar{P}_{\text{uni}}$, capturing comprehensive relationships from an item-centric view derived from both group-level and member-level interactions. 

\begin{figure}[t]
    \centering
    \begin{subfigure}[b]{0.3\linewidth}
        \includegraphics[width=\linewidth]{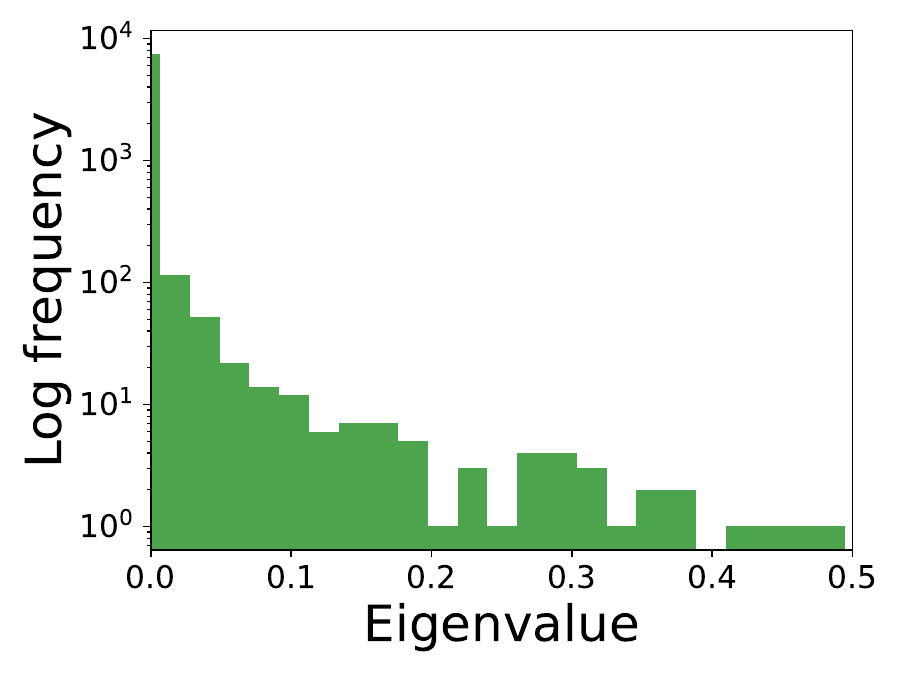}
        
        \caption{$\bar{P}_g$}
        \label{fig:L}
    \end{subfigure}
    \hfill
    \begin{subfigure}[b]{0.3\linewidth}
        \includegraphics[width=\linewidth]{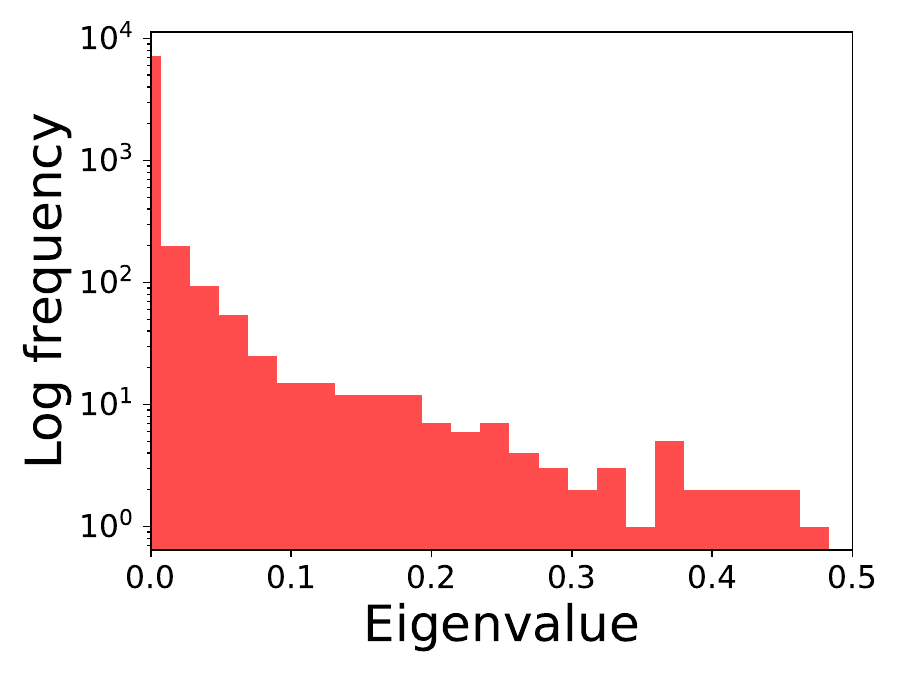}
        
        \caption{$\bar{P}_u$}
        \label{fig:I}
    \end{subfigure}
    \hfill
    \begin{subfigure}[b]{0.3\linewidth}
        \includegraphics[width=\linewidth]{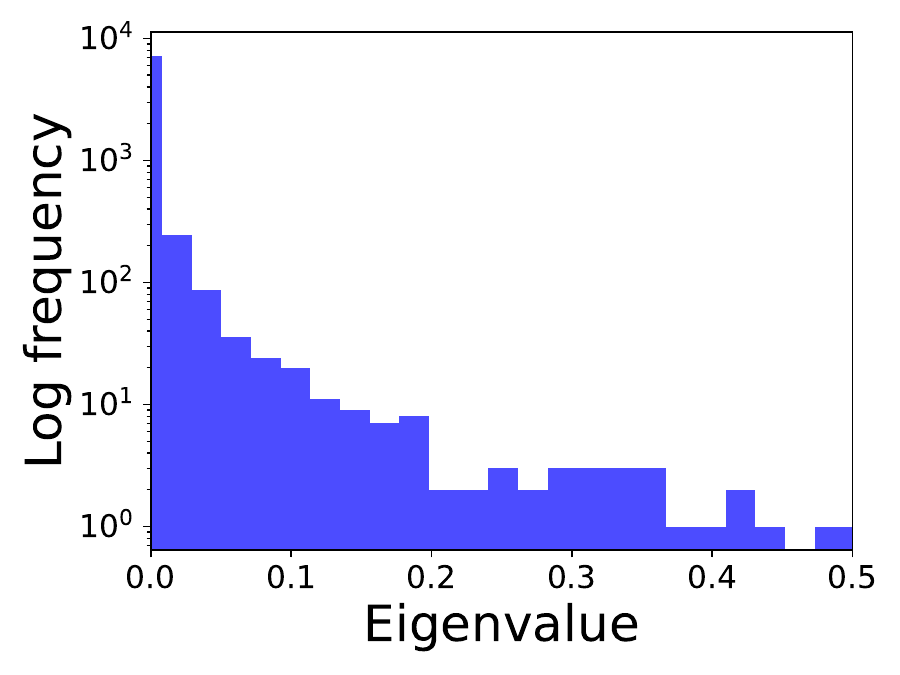}
        
        \caption{$\bar{P}_{\text{uni}}$}
        \label{fig:O}
    \end{subfigure}
    \caption{Eigenvalue distributions of the three item similarity graphs on the CAMRa2011 dataset.}
    \label{eigen_dist}
\end{figure}

\subsection{Filter Design}\label{sec 3.2.3}
We address the design principle of our multi-view GF in \textsf{Group-GF} based on the constructed item similarity graphs. To this end, we investigate the eigenvalue (\textit{i.e.}, the graph frequency) distributions of the three item similarity graphs $\bar{P}_{g}$, $\bar{P}_{u}$, and $\bar{P}_{\text{uni}}$. As illustrated in Figure \ref{eigen_dist}, the eigenvalue distributions of $\bar{P}_{g}$, $\bar{P}_{u}$, and $\bar{P}_{\text{uni}}$ tend to differ noticeably.\footnote{We empirically confirm this by measuring the Kullback-Leibler (KL) divergence: $\mathbb{KL}(p_{g}||p_{u})= 0.1010$, $\mathbb{KL}(p_{\text{uni}}||p_{g})= 0.1570$, and $\mathbb{KL}(p_{\text{uni}}||p_{u})= 0.0901$, where $p_*$ denotes the eigenvalue distribution of $\bar{P}_*$ ($*\in\{u,g,\text{uni}\}$). }   
In this context, employing the same LPF across all item similarity graphs may not optimally take advantage of the unique structural properties of each graph. Instead, we discover a {\it distinct} graph filter for each item similarity graph, ensuring that the dynamics of both groups and members are optimally captured. More specifically, based on the three item similarity graphs $\bar{P}_{u}$, $\bar{P}_{g}$, and $\bar{P}_{\text{uni}}$, we are interested in optimally discovering the polynomial graph filter for each graph, which is expressed as
\begin{equation}
    f_1(\bar{P}_{u}) = \sum_{k=1}^{K_u}{a_{k}}\bar{P}_{u}^k;  \\
    f_2(\bar{P}_{g}) = \sum_{k=1}^{K_g}{b_{k}}\bar{P}_{g}^k;  \\
    f_3(\bar{P}_{\text{uni}}) = \sum_{k=1}^{K_{\text{uni}}}{c_{k}}\bar{P}_{\text{uni}}^k,
    \label{gf_filter}
\end{equation}
where $f_1(\bar{P}_{u})$, $f_2 (\bar{P}_g)$, and $f_3 (\bar{P}_\text{uni})$ are the polynomial graph filters for $\bar{P}_{u}$, $\bar{P}_g$, and $\bar{P}_\text{uni}$, respectively; $K_u$, $K_g$, and $K_{\text{uni}}$ are the maximum order of polynomial filters for $\bar{P}_u$, $\bar{P}_g$, and $\bar{P}_\text{uni}$, respectively; and $a_k$, $b_k$, and $c_k$ are polynomial coefficients. As an example, we can use a polynomial graph filter $f_2(\bar{P}_g)=2\bar{P}_{g}-\bar{P}^2_{g}$ to find the graph filter whose frequency response function is $h(\lambda) = 1-\lambda^2$ \cite{park2024turbo,park2025criteria}. 

Due to the fact that the member-level and group-level interactions behave differently depending on group recommendation scenarios \cite{guo2021hierarchical, cao2018attentive, jia2021hypergraph}, \textsf{Group-GF} optimally aggregates the three graph filters $f_1(\bar{P}_{u}), f_2(\bar{P}_{g}),$ and $f_3(\bar{P}_{\text{uni}})$ according to
\begin{equation}
\label{group-gf}
    \mathbf{s}_g = \mathbf{r}_g((1-\alpha-\beta)f_1(\bar{P}_{u}) + \alpha f_2(\bar{P}_{g}) + \beta f_3(\bar{P}_{\text{uni}})),
\end{equation}
where ${\mathbf{r}}_{g}$ is the $g$-th row of $R_g$, representing the graph signal for group $g$; $\mathbf{s}_g$ is the predicted score for group $g$; and $\alpha$ and $\beta$ are the parameters that balance among the three types of polynomial graph filters. The final prediction optimally combines the preferences propagated through member-level, group-level, and unified interactions via hyperparameter tuning on the validation set. 



Next, let us turn to providing a theoretical foundation for \textsf{Group-GF}, which bridges between multi-view GF and optimization via smoothness regularization. We consider smoothness regularization in the context of group recommendation, because the preferences of individual members and the preferences of the belonging group as a whole are interconnected within the graph structure. The smoothness regularization assures that similar nodes (\textit{i.e.}, items) in each graph have similar preference scores, leading to more accurate recommendations.
\begin{theorem}
\small{
\label{theorem1}
Let $L_u = I - \bar{P}_u$, $L_g = I - \bar{P}_g$, and $L_{\text{uni}} = I - \bar{P}_{\text{uni}}$ be the graph Laplacians corresponding to $\bar{P}_u$, $\bar{P}_g$, and $\bar{P}_{\text{uni}}$, respectively. Then, the predicted group preference scores $\mathbf{s}_g$ in Eq. \eqref{group-gf} are an approximate solution to the following optimization problem:
\begin{equation}
\label{smootheness_eq}
\begin{split}
\mathbf{s}_g \approx \arg\min_{\mathbf{s}} \Biggl\{ & \|\mathbf{s} - \mathbf{r}_g\|^2 + \lambda \mathbf{s}^\top \left( (1 - \alpha - \beta) L_u + \alpha L_g + \beta L_{\text{uni}} \right) \mathbf{s} \Biggr\},
\end{split}
\end{equation}
where $\lambda > 0$ is a regularization parameter and $0<\alpha,\beta<1$ are balancing parameters.
}
\end{theorem} 
The proof of this theorem is available at \url{https://github.com/chaehyun1/Group-GF}. 
From Theorem \ref{theorem1}, it is shown that the predicted group preference scores $\mathbf{s}_g$ in Eq. \eqref{group-gf} are indeed an approximate solution to the optimization problem in Eq. \eqref{smootheness_eq} that contains smoothness over the member-level, group-level, and unified interactions. Here, without the regularization term (\textit{i.e.}, $\lambda = 0$), the predicted group preference scores $\mathbf{s}_g$ are equal to the observed group-level interactions $\mathbf{r}_g$. This implies that the \textsf{Group-GF} method does not integrate any information from the three item similarity graphs, thereby ignoring the relationships captured by the member-level, group-level, and unified interactions. This theoretical finding enhances the interpretability of the \textsf{Group-GF} method by linking group preference propagation with smoothness regularization. In other words, this interpretation provides a clearer understanding of how preferences are identified from the three different interactions.

\section{Experimental Evaluation}

\definecolor{lightgray}{gray}{0.9}
\begin{table}[t]
\footnotesize
  \captionsetup{skip=1pt}
  \caption{The statistics of the three real-world benchmark datasets having both group--item and member--item interactions.}
  \begin{tabular}{@{}ccccccl@{}}
    \toprule
    Dataset & \# Members & \# Items & \# Groups & \# M-I inter. & \# G-I inter. \\ 
    \midrule
    CAMRa2011 & 602 & 7,710 & 290 & 116,344 & 145,068 \\
    Mafengwo & 5,275 & 1,513 & 995 & 39,761 & 3,595 \\
    Douban  & 3,481 & 43,916 & 1,162 & 829,229 & 9,245 \\
  \bottomrule
\end{tabular}
\label{table:datasets}
\end{table}

\label{sec4}

\subsection{Experimental Settings}
\noindent\textbf{Datasets.} We adopt three widely used benchmark datasets on group recommendations \cite{wu2023consrec, cao2018attentive, jia2021hypergraph, yin2019social}, including CAMRa2011, Mafengwo, and Douban. Table \ref{table:datasets} summarizes the statistics of the three datasets. We note that the size of group recommendation datasets are inherently smaller in scale than that of canonical (user) recommendation datasets, reflecting their focus on collective interactions rather than individual-level data.

\noindent\textbf{Competitors.} We compare \textsf{Group-GF} with eight state-of-the-art group recommendation methods, including DNN-based (NCF \cite{he2017neural}), attentive aggregation-based (AGREE \cite{cao2018attentive}), hypergraph-based (HyperGroup \cite{guo2021hierarchical}, HCR \cite{jia2021hypergraph}, and ConsRec \cite{wu2023consrec}), and SSL-based (GroupIM \cite{sankar2020groupim}, S$^2$-HHGR \cite{zhang2021double}, and CubeRec \cite{chen2022thinking}) methods.

\noindent\textbf{Evaluation protocols.} Following the evaluation protocol outlined in \cite{jia2021hypergraph, wu2023consrec}, we adopt performance metrics such as the hit ratio (HR@$k$) and the normalized discounted cumulative gain (NDCG@$k$), where $k$ is set to 10 by default, due to the space limitation. 

\noindent\textbf{Implementation details.} In our experiments, rather than exhaustively searching for polynomial coefficients, we use the polynomial graph filters presented in \cite{park2024turbo} ({\it i.e.}, pre-defined polynomial coefficients in Eq. (\ref{gf_filter})) for efficient filter search, because it was empirically confirmed that exhaustive search leads to only negligible gains in recommendation accuracy. \textsf{Group-GF} is basically training-free; however, similarly as in training-based recommendation methods, it adjusts hyperparameters on the validation set. All experiments are carried out with the same device: Intel (R) 12-Core (TM) i7-9700K CPUs @ 3.60 GHz and GPU of NVIDIA GeForce RTX A6000.

\noindent\textbf{Additional experiments.} Further experimental results, including 1) scalability, 2) analysis on different group sizes, 3) comparison with canonical recommendation methods, and 4) sensitivity analysis, as well as the reproducibility code, can be found at \url{https://github.com/chaehyun1/Group-GF}.

\begin{table}[t]
\footnotesize 
\centering
\caption{Runtime comparison. The best and second-best performers are highlighted in bold and underline, respectively. Here, `OOM' denotes an out-of-memory issue.}
\label{runtime_table}
\begin{tabular}{lccccc} 
\hline
 & AGREE & GroupIM & CubeRec & ConsRec & \textbf{Group-GF} \\ \hline
CAMRa2011 & 1h41m6s & \underline{1m34s} & 4h16m4s & 12m19s & \textbf{6.71s} \\ 
Mafengwo  & 24m1s  & \underline{57.13s} & 5m45s   & 14m57s & \textbf{1.55s} \\ 
Douban  & 12h1m36s  & OOM & 1h28m51s & \underline{1h21m1s} & \textbf{3m26s} \\ 
\hline
Training  & \textcolor{blue}{\cmark} & \textcolor{blue}{\cmark} & \textcolor{blue}{\cmark} & \textcolor{blue}{\cmark} & \textcolor{purple}{\xmark} \\ \hline
\end{tabular}
\end{table}

\begin{table}[t!] 
\vspace{-3mm}
\footnotesize
\centering
\caption{Recommendation accuracy (HR@10 and NDCG@10). The best and second-best performers are highlighted in bold and underline, respectively.}

\label{main_results}
\begin{tabular}{lcccccc}
\toprule
 & \multicolumn{2}{c}{\textbf{CAMRa2011}} & \multicolumn{2}{c}{\textbf{Mafengwo}} & \multicolumn{2}{c}{\textbf{Douban}} \\
\cmidrule(r){2-3} \cmidrule(r){4-5} \cmidrule(r){6-7}
 & HR & NDCG & HR & NDCG& HR & NDCG \\
\midrule
NCF & 0.7693 & 0.4448 & 0.6269 & 0.4141 &0.4132 & 0.2446 \\
AGREE & 0.7789 & 0.4530 & 0.6321 & 0.4203 & 0.7701 & \underline{0.6859} \\
HyperGroup & 0.7986 & 0.4538 & 0.6482 & 0.5018 & \underline{0.7757} & 0.5318 \\
HCR & 0.7821 & 0.4670 & 0.8503 & 0.6852 & 0.6537 & 0.6441 \\
GroupIM & \underline{0.8407} & 0.4914 & 0.8161 & 0.6330 & \multicolumn{2}{c}{------ OOM ------} \\
S$^2$-HHGR & 0.7903 & 0.4453 & 0.7779 & 0.7391 & 0.7180 & 0.4565 \\
CubeRec & 0.8207 & 0.4935 & 0.9025 & 0.7708 & 0.6992 & 0.4240 \\
ConsRec & 0.8248 & \underline{0.4945} & \underline{0.9156} & \underline{0.7794} & 0.7403 & 0.6621 \\
\rowcolor{lightgray}\textbf{Group-GF} & \textbf{0.9552} & \textbf{0.5030} & \textbf{0.9266} & \textbf{0.8451} & \textbf{0.9028} & \textbf{0.7291} \\
\bottomrule
\end{tabular}
\end{table}

\vspace{-2mm}

\subsection{Runtime Analysis}
Table \ref{runtime_table} presents the runtime of \textsf{Group-GF} compared to the four DNN-based competitors that perform well (AGREE, GroupIM, CubeRec, and ConsRec) on the three benchmark datasets. Here, the runtime indicates the training duration for DNN-based methods, while referring to the processing time for \textsf{Group-GF}. 
\textsf{Group-GF} exhibits its remarkable computational efficiency, achieving up to \underline{$\times36.85$ faster runtime} than that of the second-best performer ({\it i.e.}, GroupIM) on the Mafengwo dataset.
This is because \textsf{Group-GF} operates solely on straightforward matrix operations, eliminating the need for a costly training process to learn patterns in group-level and member-level interactions. A similar tendency can be observed in both the CAMRa2011 and Douban datasets.

\subsection{Recommendation Accuracy}
The performance of \textsf{Group-GF} and all competitors is summarized in Table \ref{main_results}. Our key observations are made as follows:
\begin{enumerate}[label=(\roman*)]
    \item In spite of not requiring any training, \textsf{Group-GF} consistently outperforms state-of-the-art group recommendation methods across all datasets and metrics. Specifically, on the Douban dataset, it achieves up to 16.4\% higher HR@10 than that of the best competitor. This is due to \textsf{Group-GF}'s ability to effectively integrate member and group information while capturing complex member-level, group-level, and unified interactions.
    \item In contrast to early-developed methods such as AGREE, techniques based on hypergraph structures and SSL show comparatively competing performance. This implies the importance of modeling the complex interactions between each group and its belonging members in capturing group preferences.
\end{enumerate}
\vspace{-2mm}

\begin{table}[t]
\small
\centering
\caption{Performance comparison among \textsf{Group-GF} and its four variants in terms of NDCG@10.}

\label{ablation_table_modified_ndcg}
\begin{tabular}{lccc}
\hline
\textbf{} & \textbf{CAMRa2011} & \textbf{Mafengwo} & \textbf{Douban}\\  \hline
\textsf{Group-GF} & \textbf{0.5030} & \textbf{0.8451} & \textbf{0.7291}\\ 
\textsf{Group-GF-m} & 0.4769 & 0.8376 & 0.6104\\ 
\textsf{Group-GF-g} & 0.5003 & 0.8367 & 0.7272\\ 
\textsf{Group-GF-uni} & 0.4999 & 0.8272& 0.6665\\ 
\textsf{Group-GF-a} & 0.5025 & 0.8446& 0.6728\\ 
\hline
\end{tabular}
\end{table}

\subsection{Ablation Study}

To assess the contribution of each component in \textsf{Group-GF}, we perform an extensive ablation study alongside four variants, each based on different sources accounting for \textsf{Group-GF}: 1) \textsf{Group-GF-m}: excludes the member-level component $\bar{P}_{u}$ in Eq. (\ref{adj_graph_ug}); 2) \textsf{Group-GF-g}: excludes the group-level component $\bar{P}_{g}$ in Eq. (\ref{adj_graph_ug}); 3) \textsf{Group-GF-uni}: excludes the unified item similarity graph $\bar{P}_{\text{uni}}$ in Eq. (\ref{adj_graph_uni}); 4) \textsf{Group-GF-a}: removes the member--group relation matrix $M$ in Eq. (\ref{aug_inters}).
 The performance comparison among the original \textsf{Group-GF} and its four variants is summarized in Table \ref{ablation_table_modified_ndcg} {\it w.r.t.} the NDCG@10. Our findings are as follows:

\begin{enumerate}[label=(\roman*)]

    \item \textsf{Group-GF} consistently outperforms all four variants, emphasizing the vital role of each component in ensuring accurate group recommendations.

    \item \textsf{Group-GF-m} exhibits the largest decline on CAMRa2011 and Douban, while \textsf{Group-GF-uni} on Mafengwo. This implies that the importance of types of interactions varies depending on the dataset characteristics.
        

    \item The absence of $\bar{P}_{\text{uni}}$ leads to significant performance drops consistently across all the datasets, highlighting the importance of jointly incorporating group-level and member-level interactions in GF. A more nuanced reflection of all possible interactions between members and groups can facilitate more accurate group recommendations.

    \item The removal of $M$ from Eq. \eqref{aug_inters} also results in a non-negligible performance decrease on all the datasets. This observation indicates that reflecting group--member relationship information in GF is crucial for achieving state-of-the-art recommendation performance.
\end{enumerate}

\section{Conclusions and Future Work}
We addressed an unexplored yet important challenge of designing GF methods for group recommendations.
To this end, we proposed \textsf{Group-GF}, an innovative group recommendation method via training-free multi-view GF, which is capable of jointly leveraging all available interactions. In \textsf{Group-GF}, we showed 1) how to construct two augmented item similarity graphs and one unified item similarity graph, which manifest different viewpoints for GF, and 2) how to efficiently and optimally discover each of distinct polynomial graph filters and aggregate them.
Through systematic evaluations and analyses, we demonstrated (a) the remarkable computational efficiency of \textsf{Group-GF}, (b) the superior recommendation accuracy over state-of-the-art methods in various circumstances, and (c) the theoretical finding that supports the model's performance and interpretability.
As the three similarity graphs $\bar{P}_{u}$, $\bar{P}_{g}$, and $\bar{P}_{\text{uni}}$ need to be loaded into memory, which is memory-demanding, potential avenues of our future research include the design of a memory-efficient and scalable GF method for large-scale group recommendation scenarios.

\section*{Acknowledgments}
This work was supported by the National Research Foundation of Korea (NRF), Republic of Korea Grant by the Korean Government through MSIT under Grants RS-2021-NR059723 and RS-2023-00220762 and by the Institute of Information and Communications Technology Planning and Evaluation (IITP), Republic of Korea Grant by the Korean Government through MSIT (6G Post-MAC–POsitioning and Spectrum-Aware intelligenT MAC for Computing and Communication Convergence) under Grant 2021-0-00347.
\vspace{-1mm}

\bibliographystyle{ACM-Reference-Format}

\bibliography{citation_list}


\begin{thebibliography}{21}


\ifx \showCODEN    \undefined \def \showCODEN     #1{\unskip}     \fi
\ifx \showISBNx    \undefined \def \showISBNx     #1{\unskip}     \fi
\ifx \showISBNxiii \undefined \def \showISBNxiii  #1{\unskip}     \fi
\ifx \showISSN     \undefined \def \showISSN      #1{\unskip}     \fi
\ifx \showLCCN     \undefined \def \showLCCN      #1{\unskip}     \fi
\ifx \shownote     \undefined \def \shownote      #1{#1}          \fi
\ifx \showarticletitle \undefined \def \showarticletitle #1{#1}   \fi
\ifx \showURL      \undefined \def \showURL       {\relax}        \fi
\providecommand\bibfield[2]{#2}
\providecommand\bibinfo[2]{#2}
\providecommand\natexlab[1]{#1}
\providecommand\showeprint[2][]{arXiv:#2}

\bibitem[Baltrunas et~al\mbox{.}(2010)]%
        {baltrunas2010group}
\bibfield{author}{\bibinfo{person}{Linas Baltrunas}, \bibinfo{person}{Tadas Makcinskas}, {and} \bibinfo{person}{Francesco Ricci}.} \bibinfo{year}{2010}\natexlab{}.
\newblock \showarticletitle{Group recommendations with rank aggregation and collaborative filtering}. In \bibinfo{booktitle}{\emph{RecSys}}. \bibinfo{pages}{119--126}.
\newblock


\bibitem[Cao et~al\mbox{.}(2018)]%
        {cao2018attentive}
\bibfield{author}{\bibinfo{person}{Da Cao}, \bibinfo{person}{Xiangnan He}, \bibinfo{person}{Lianhai Miao}, \bibinfo{person}{Yahui An}, \bibinfo{person}{Chao Yang}, {and} \bibinfo{person}{Richang Hong}.} \bibinfo{year}{2018}\natexlab{}.
\newblock \showarticletitle{Attentive group recommendation}. In \bibinfo{booktitle}{\emph{SIGIR}}. \bibinfo{pages}{645--654}.
\newblock


\bibitem[Chen et~al\mbox{.}(2022)]%
        {chen2022thinking}
\bibfield{author}{\bibinfo{person}{Tong Chen}, \bibinfo{person}{Hongzhi Yin}, \bibinfo{person}{Jing Long}, \bibinfo{person}{Quoc Viet~Hung Nguyen}, \bibinfo{person}{Yang Wang}, {and} \bibinfo{person}{Meng Wang}.} \bibinfo{year}{2022}\natexlab{}.
\newblock \showarticletitle{Thinking inside the box: {L}earning hypercube representations for group recommendation}. In \bibinfo{booktitle}{\emph{SIGIR}}. \bibinfo{pages}{1664--1673}.
\newblock


\bibitem[Choi et~al\mbox{.}(2023)]%
        {choi2023blurring}
\bibfield{author}{\bibinfo{person}{Jeongwhan Choi}, \bibinfo{person}{Seoyoung Hong}, \bibinfo{person}{Noseong Park}, {and} \bibinfo{person}{Sung-Bae Cho}.} \bibinfo{year}{2023}\natexlab{}.
\newblock \showarticletitle{Blurring-sharpening process models for collaborative filtering}. In \bibinfo{booktitle}{\emph{SIGIR}}. \bibinfo{pages}{1096--1106}.
\newblock


\bibitem[Guo et~al\mbox{.}(2021)]%
        {guo2021hierarchical}
\bibfield{author}{\bibinfo{person}{Lei Guo}, \bibinfo{person}{Hongzhi Yin}, \bibinfo{person}{Tong Chen}, \bibinfo{person}{Xiangliang Zhang}, {and} \bibinfo{person}{Kai Zheng}.} \bibinfo{year}{2021}\natexlab{}.
\newblock \showarticletitle{Hierarchical hyperedge embedding-based representation learning for group recommendation}.
\newblock \bibinfo{journal}{\emph{TOIS}} \bibinfo{volume}{40}, \bibinfo{number}{1} (\bibinfo{year}{2021}), \bibinfo{pages}{1--27}.
\newblock


\bibitem[He et~al\mbox{.}(2020)]%
        {he2020lightgcn}
\bibfield{author}{\bibinfo{person}{Xiangnan He}, \bibinfo{person}{Kuan Deng}, \bibinfo{person}{Xiang Wang}, \bibinfo{person}{Yan Li}, \bibinfo{person}{Yongdong Zhang}, {and} \bibinfo{person}{Meng Wang}.} \bibinfo{year}{2020}\natexlab{}.
\newblock \showarticletitle{Lightgcn: Simplifying and powering graph convolution network for recommendation}. In \bibinfo{booktitle}{\emph{SIGIR}}. \bibinfo{pages}{639--648}.
\newblock


\bibitem[He et~al\mbox{.}(2017)]%
        {he2017neural}
\bibfield{author}{\bibinfo{person}{Xiangnan He}, \bibinfo{person}{Lizi Liao}, \bibinfo{person}{Hanwang Zhang}, \bibinfo{person}{Liqiang Nie}, \bibinfo{person}{Xia Hu}, {and} \bibinfo{person}{Tat-Seng Chua}.} \bibinfo{year}{2017}\natexlab{}.
\newblock \showarticletitle{Neural collaborative filtering}. In \bibinfo{booktitle}{\emph{WWW}}. \bibinfo{pages}{173--182}.
\newblock


\bibitem[Hu et~al\mbox{.}(2014)]%
        {hu2014deep}
\bibfield{author}{\bibinfo{person}{Liang Hu}, \bibinfo{person}{Jian Cao}, \bibinfo{person}{Guandong Xu}, \bibinfo{person}{Longbing Cao}, \bibinfo{person}{Zhiping Gu}, {and} \bibinfo{person}{Wei Cao}.} \bibinfo{year}{2014}\natexlab{}.
\newblock \showarticletitle{Deep modeling of group preferences for group-based recommendation}. In \bibinfo{booktitle}{\emph{AAAI}}. \bibinfo{pages}{1861–1867}.
\newblock


\bibitem[Jia et~al\mbox{.}(2021)]%
        {jia2021hypergraph}
\bibfield{author}{\bibinfo{person}{Renqi Jia}, \bibinfo{person}{Xiaofei Zhou}, \bibinfo{person}{Linhua Dong}, {and} \bibinfo{person}{Shirui Pan}.} \bibinfo{year}{2021}\natexlab{}.
\newblock \showarticletitle{Hypergraph convolutional network for group recommendation}. In \bibinfo{booktitle}{\emph{ICDM}}. \bibinfo{pages}{260--269}.
\newblock


\bibitem[Liu et~al\mbox{.}(2023)]%
        {liu2023personalized}
\bibfield{author}{\bibinfo{person}{Jiahao Liu}, \bibinfo{person}{Dongsheng Li}, \bibinfo{person}{Hansu Gu}, \bibinfo{person}{Tun Lu}, \bibinfo{person}{Peng Zhang}, \bibinfo{person}{Li Shang}, {and} \bibinfo{person}{Ning Gu}.} \bibinfo{year}{2023}\natexlab{}.
\newblock \showarticletitle{Personalized graph signal processing for collaborative filtering}. In \bibinfo{booktitle}{\emph{WWW}}. \bibinfo{pages}{1264--1272}.
\newblock


\bibitem[Park et~al\mbox{.}(2024)]%
        {park2024turbo}
\bibfield{author}{\bibinfo{person}{Jin-Duk Park}, \bibinfo{person}{Yong-Min Shin}, {and} \bibinfo{person}{Won-Yong Shin}.} \bibinfo{year}{2024}\natexlab{}.
\newblock \showarticletitle{Turbo-CF: Matrix decomposition-free graph filtering for fast recommendation}. In \bibinfo{booktitle}{\emph{SIGIR}}. \bibinfo{pages}{2672--2676}.
\newblock


\bibitem[Park et~al\mbox{.}(2025)]%
        {park2025criteria}
\bibfield{author}{\bibinfo{person}{Jin-Duk Park}, \bibinfo{person}{Jaemin Yoo}, {and} \bibinfo{person}{Won-Yong Shin}.} \bibinfo{year}{2025}\natexlab{}.
\newblock \showarticletitle{Criteria-aware graph filtering: Extremely fast yet accurate multi-criteria recommendation}. In \bibinfo{booktitle}{\emph{WWW}}. \bibinfo{pages}{(to appear)}.
\newblock


\bibitem[Sankar et~al\mbox{.}(2020)]%
        {sankar2020groupim}
\bibfield{author}{\bibinfo{person}{Aravind Sankar}, \bibinfo{person}{Yanhong Wu}, \bibinfo{person}{Yuhang Wu}, \bibinfo{person}{Wei Zhang}, \bibinfo{person}{Hao Yang}, {and} \bibinfo{person}{Hari Sundaram}.} \bibinfo{year}{2020}\natexlab{}.
\newblock \showarticletitle{Group{IM}: A mutual information maximization framework for neural group recommendation}. In \bibinfo{booktitle}{\emph{SIGIR}}. \bibinfo{pages}{1279--1288}.
\newblock


\bibitem[Shen et~al\mbox{.}(2021)]%
        {shen2021powerful}
\bibfield{author}{\bibinfo{person}{Yifei Shen}, \bibinfo{person}{Yongji Wu}, \bibinfo{person}{Yao Zhang}, \bibinfo{person}{Caihua Shan}, \bibinfo{person}{Jun Zhang}, \bibinfo{person}{B~Khaled Letaief}, {and} \bibinfo{person}{Dongsheng Li}.} \bibinfo{year}{2021}\natexlab{}.
\newblock \showarticletitle{How powerful is graph convolution for recommendation?}. In \bibinfo{booktitle}{\emph{CIKM}}. \bibinfo{pages}{1619--1629}.
\newblock


\bibitem[Shuman et~al\mbox{.}(2013)]%
        {shuman2013emerging}
\bibfield{author}{\bibinfo{person}{David~I Shuman}, \bibinfo{person}{Sunil~K Narang}, \bibinfo{person}{Pascal Frossard}, \bibinfo{person}{Antonio Ortega}, {and} \bibinfo{person}{Pierre Vandergheynst}.} \bibinfo{year}{2013}\natexlab{}.
\newblock \showarticletitle{The emerging field of signal processing on graphs: Extending high-dimensional data analysis to networks and other irregular domains}.
\newblock \bibinfo{journal}{\emph{IEEE Signal Processing Magazine}} \bibinfo{volume}{30}, \bibinfo{number}{3} (\bibinfo{year}{2013}), \bibinfo{pages}{83--98}.
\newblock


\bibitem[Wu et~al\mbox{.}(2023)]%
        {wu2023consrec}
\bibfield{author}{\bibinfo{person}{Xixi Wu}, \bibinfo{person}{Yun Xiong}, \bibinfo{person}{Yao Zhang}, \bibinfo{person}{Yizhu Jiao}, \bibinfo{person}{Jiawei Zhang}, \bibinfo{person}{Yangyong Zhu}, {and} \bibinfo{person}{Philip~S Yu}.} \bibinfo{year}{2023}\natexlab{}.
\newblock \showarticletitle{Cons{R}ec: Learning consensus behind interactions for group recommendation}. In \bibinfo{booktitle}{\emph{WWW}}. \bibinfo{pages}{240--250}.
\newblock


\bibitem[Xia et~al\mbox{.}(2022)]%
        {xia2022fire}
\bibfield{author}{\bibinfo{person}{Jiafeng Xia}, \bibinfo{person}{Dongsheng Li}, \bibinfo{person}{Hansu Gu}, \bibinfo{person}{Jiahao Liu}, \bibinfo{person}{Tun Lu}, {and} \bibinfo{person}{Ning Gu}.} \bibinfo{year}{2022}\natexlab{}.
\newblock \showarticletitle{FIRE: Fast incremental recommendation with graph signal processing}. In \bibinfo{booktitle}{\emph{WWW}}. \bibinfo{pages}{2360--2369}.
\newblock


\bibitem[Yadati et~al\mbox{.}(2019)]%
        {yadati2019hypergcn}
\bibfield{author}{\bibinfo{person}{Naganand Yadati}, \bibinfo{person}{Madhav Nimishakavi}, \bibinfo{person}{Prateek Yadav}, \bibinfo{person}{Vikram Nitin}, \bibinfo{person}{Anand Loauis}, {and} \bibinfo{person}{Partha Talukdar}.} \bibinfo{year}{2019}\natexlab{}.
\newblock \showarticletitle{Hyper{GCN}: A new method for training graph convolutional networks on hypergraphs}.
\newblock \bibinfo{journal}{\emph{NeurIPS}}, \bibinfo{pages}{1511–1522}.
\newblock


\bibitem[Yang et~al\mbox{.}(2023)]%
        {yang2023group}
\bibfield{author}{\bibinfo{person}{Mingdai Yang}, \bibinfo{person}{Zhiwei Liu}, \bibinfo{person}{Liangwei Yang}, \bibinfo{person}{Xiaolong Liu}, \bibinfo{person}{Chen Wang}, \bibinfo{person}{Hao Peng}, {and} \bibinfo{person}{Philip~S Yu}.} \bibinfo{year}{2023}\natexlab{}.
\newblock \showarticletitle{Group identification via transitional hypergraph convolution with cross-view self-supervised learning}. In \bibinfo{booktitle}{\emph{CIKM}}. \bibinfo{pages}{2969--2979}.
\newblock


\bibitem[Yin et~al\mbox{.}(2019)]%
        {yin2019social}
\bibfield{author}{\bibinfo{person}{Hongzhi Yin}, \bibinfo{person}{Qinyong Wang}, \bibinfo{person}{Kai Zheng}, \bibinfo{person}{Zhixu Li}, \bibinfo{person}{Jiali Yang}, {and} \bibinfo{person}{Xiaofang Zhou}.} \bibinfo{year}{2019}\natexlab{}.
\newblock \showarticletitle{Social influence-based group representation learning for group recommendation}. In \bibinfo{booktitle}{\emph{ICDE}}. \bibinfo{pages}{566--577}.
\newblock


\bibitem[Zhang et~al\mbox{.}(2021)]%
        {zhang2021double}
\bibfield{author}{\bibinfo{person}{Junwei Zhang}, \bibinfo{person}{Min Gao}, \bibinfo{person}{Junliang Yu}, \bibinfo{person}{Lei Guo}, \bibinfo{person}{Jundong Li}, {and} \bibinfo{person}{Hongzhi Yin}.} \bibinfo{year}{2021}\natexlab{}.
\newblock \showarticletitle{Double-scale self-supervised hypergraph learning for group recommendation}. In \bibinfo{booktitle}{\emph{CIKM}}. \bibinfo{pages}{2557--2567}.
\newblock


\end{thebibliography}

\end{document}